\documentclass[journal]{IEEEtran}
\ifCLASSINFOpdf

\else

\fi
\usepackage[cmex10]{amsmath}
\usepackage{amssymb}
\usepackage{cite}
\usepackage{graphicx}
\usepackage{array,color}
\usepackage{amsmath}
\usepackage{stfloats}
\usepackage{graphicx}
\usepackage{subfigure}
\usepackage{tabularx}
\usepackage{epsfig,epsf,color,balance,cite}
\usepackage{algorithmic}
\usepackage{algorithm}
\usepackage{bm}
\usepackage{textcomp}

\begin{document}

\title{ Reconfigurable Intelligent Surfaces for 6G Systems:   Principles, Applications, and Research Directions}
\author{ Cunhua Pan, Hong Ren, Kezhi Wang, Jonas Florentin Kolb,  Maged Elkashlan, Ming Chen, Marco Di Renzo, \IEEEmembership{Fellow, IEEE}, Yang Hao, \IEEEmembership{Fellow, IEEE}, Jiangzhou Wang,  \IEEEmembership{Fellow, IEEE}, A. Lee Swindlehurst, \IEEEmembership{Fellow, IEEE},  Xiaohu You, \IEEEmembership{Fellow, IEEE} and Lajos Hanzo, \IEEEmembership{Fellow, IEEE}

\thanks{C. Pan, J. Kolb, M. Elkashlan and Y. Hao are with  Queen Mary University of London, U.K. (Email:\{c.pan, j.f.kolb, maged.elkashlan, y.hao\}@qmul.ac.uk). H. Ren, M. Chen, and X. You  are with  Southeast University. (Email: {hren, chenming, xhyu}@seu.edu.cn). Kezhi Wang is with Northumbria University, UK.  (Email: kezhi.wang@northumbria.ac.uk). M. Di Renzo
is with Universite Paris-Saclay,  France. (Email: marco.direnzo@centralesupelec.fr). Jiangzhou Wang is with University of Kent, UK. (Email: J.Z.Wang@kent.ac.uk). A. L. Swindlehurst is with University of California at Irvine,  USA (Email: swindle@uci.edu). Lajos Hanzo is with University of Southampton, Southampton,  U.K. (Email: lh@ecs.soton.ac.uk).}

\thanks{This work has been submitted to the IEEE for possible publication. Copyright may be transferred without notice,
after which this version may no longer be accessible.}
}


\maketitle

\begin{abstract}
 Reconfigurable intelligent surfaces (RISs) or intelligent reflecting surfaces (IRSs),  are regarded as one of the most promising and revolutionizing techniques for enhancing the spectrum and/or energy efficiency of wireless systems. These devices are capable of  reconfiguring the wireless propagation environment by carefully tuning the phase shifts of a large number of low-cost passive reflecting elements.  In this article, we aim for answering four fundmental questions: 1) Why do we need RISs? 2) What is an RIS? 3) What are RIS's applications? 4) What are  the relevant challenges and future research directions? In response, eight promising research directions are pointed out.
\end{abstract}

\IEEEpeerreviewmaketitle
\section{Why Do We Need RIS?}

Although   fifth-generation (5G) wireless networks are being rolled out worldwide, the key physical layer technology therein is massive multiple-input multiple-output (MIMO) operating in the sub-6 GHz bands,  while millimeter wave (mmWave) communication, originally envisioned as one of three pivotal technologies in 5G networks, has not been widely adopted.  The key impediments of  mmWave communication include its sensitivity to blockages, limited coverage, and severe path loss. However, some innovative applications, such as immersive virtual reality, high-fidelity holographic projections, digital twins, connected robotics and autonomous systems, industrial internet of things, intelligent transportation system   and brain-computer interfaces, are expected to be supported by 6G-and-beyond communications \cite{Tariq2020}. These applications entail high quality-of-service (QoS) requirements such as extremely high data rates, ultra-high reliability, and ultra-low latency, which cannot be readily supported by the existing systems. Given the large amount of available bandwidth at higher frequencies, communications in the mmWave and even Terahertz bands will be an inevitable trend. The array gain  of massive MIMO techniques at the base stations (BSs) mitigates  the  path loss at high frequencies, but  fails to solve the blockage problem. More densely deployed BSs can help eliminate blockages and fill coverage holes, but this is a costly solution both in terms of its infrastructure (and backhaul requirements) and power consumption. Hence, new cost-effective and power-efficient technologies are needed to solve these problems.

Recently,  reconfigurable intelligent surfaces (RISs), have been  envisioned as a key enabling  technology to circumvent the above-mentioned issues. RISs can be installed on large flat surfaces (e.g., walls or ceilings indoors, buildings or signage outdoors) in order to reflect radio-frequency (RF) energy around obstacles and create a virtual line-of-sight (LoS) propagation path between a mmWave source and the destination.

\section{What Is An RIS?}
 \begin{figure}[ht]
            \centering
              \includegraphics[width=0.5\textwidth]{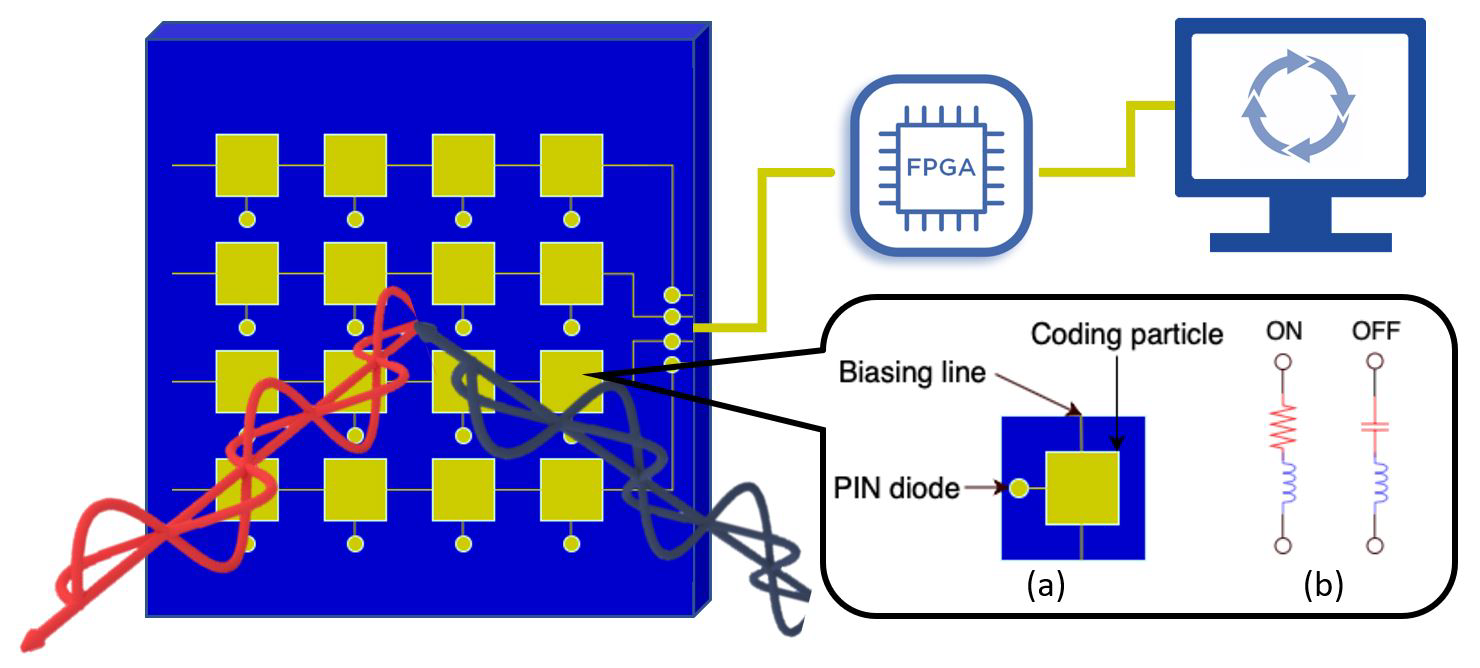}
              \caption{Architecture of an RIS.}
              \label{RISarch}
 \end{figure}
An RIS is a planar surface consisting of an array of passive reflecting elements, each of which can independently impose the required phase shift on the incoming signal \cite{wu2020intelligent,basar2019wireless}.  Based on the specific materials of the reflecting elements, the RIS can be classified into  antenna-array-based   \cite{xtan2016} and metasurface-based structures \cite{cui2014coding}.  By carefully adjusting the phase shifts of all the reflecting elements, the  reflected signals can be reconfigured to propagate towards  their desired directions. Due to  rapid developments in metamaterials, the reflection coefficient of each element can be  reconfigured in real time to adapt to the dynamically fluctuating wireless propagation environment.

 As shown in Fig. \ref{RISarch}-(a), a typical RIS architecture fabricated with metamaterials  mainly consists of a planar surface and a controller.  The planar surface may be made of a single or, in general, multiple layers. In \cite{wu2020intelligent}, for example, a three-layer planar surface is designed. The outer layer has a large number of reflecting elements printed on a dielectric substrate to directly act on the incident signals. The middle layer is a copper panel to avoid  signal/energy leakage. The last layer is a  circuit board that is used for tuning the reflection coefficients of the RIS elements,  which is operated by a smart controller such as a field-programmable gate array (FPGA). In a typical scenario envisioned for their operation,  the optimal reflection coefficients of the RIS are calculated at the BS, and then sent to the RIS's controller through a dedicated feedback link. The design of the reflection coefficients depends on the channel state information (CSI), which is only updated when the CSI changes,  on a much longer time scale than the data  symbol duration.  Fig. \ref{RISarch}-(a)  shows the structure of each reflecting element, in which     a positive-intrinsic negative
(PIN) diode is embedded. By controlling the   voltage through the biasing line, the PIN can switch between `On' and `Off' modes as shown in the equivalent circuit of  Fig. \ref{RISarch}-(b), which can realize a phase shift difference of $\pi$ in radians \cite{cui2014coding,wu2020intelligent}. To increase the number of phase shift levels, more PINs have to be integrated in each element.

RISs also have important advantages for practical implementations. For example, the  RIS reflecting elements   only passively reflect the incoming signals without any sophisticated signal processing (SP) operations that require RF transceiver hardware. Hence, compared to conventional active transmitters, RISs can operate with much lower cost in terms of hardware and power consumption \cite{wu2020intelligent,Renzo2020}. Additionally, due to the passive nature of the reflecting elements, RISs can be fabricated with light weight and limited layer thickness, hence they can be readily installed on walls, ceilings, signage,  street lamps, etc. Furthermore, an RIS naturally operates   in full-duplex (FD) mode without self-interference or introducing  thermal noise. Therefore, they achieve   higher spectral efficiency than active half-duplex (HD) relays, despite their lower signal processing complexity than that of active FD relays requiring  sophisticated self-interference cancelation. In Table \ref{tabzero}, we compare RISs to various kinds of relays, since they both serve to create   alternative transmission links. The acronyms of AF and DF in Table \ref{tabzero} refer to amplify-and-forward and decode-and-forward, respectively.

\begin{table}[!t]
\renewcommand{\arraystretch}{1.1}
\caption{  \textbf{RIS vs. Relays}}
\label{tabzero}
\centering
\begin{tabular}{l|c|c|c|c}
\hline
   & \textbf{RIS}& \textbf{AF Relay} & \textbf{DF Relay} & \textbf{FD Relay}  \\
 \hline
 With RF Chains? &No& Yes&Yes&Yes\\
 \hline
 SP Capability? &No& No&Yes&Yes\\
 \hline
 Noise?&No & Yes&Yes&Yes\\
 \hline
 Duplex &Full& Half&Half& Full\\
 \hline
 Hardware cost& Low&Median& High& Very high\\
\hline
Power Consumption& Low &Median& High& Very high\\
\hline
\end{tabular}
\end{table}

\section{What Are The RIS's Applications?}
\begin{figure*}[htbp]

\begin{center}

\includegraphics[width=0.98\textwidth]{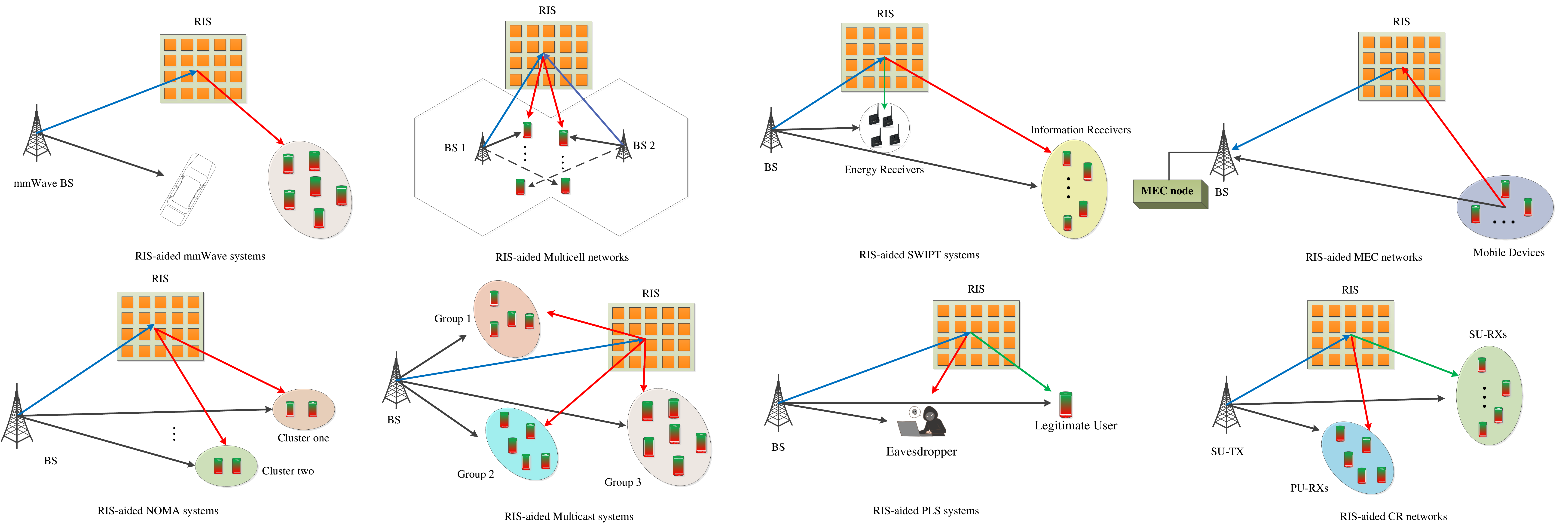}

\end{center}

\caption{Typical applications of RIS in various emerging Sub-6 GHz systems.}

\label{varioussystems}

\end{figure*}
By judiciously tuning the phase shifts of the reflecting elements of the RIS, the reflected signals can be constructively superimposed with those from   the direct paths for enhancing the desired signal power, or destructively combined for mitigating   deleterious effects of  multiuser interference. Hence, RISs provide additional degrees of freedom  to further improve the system performance. In the following, we list some typical RIS applications  in various emerging  systems.

\subsection{RIS-aided mmWave Systems} 

MmWave techniques have the potential of supporting high data rates given their high bandwidth. However, communication at mmWave frequency also has some drawbacks, such as  its severe pathloss. Fortunately, this can be mitigated by its huge array gain provided by a large antenna array within a compact space, given its short wavelength. Another impediment is that it is vulnerable to blockages by cars, pedestrians, and trees. The readily penetration loss is  also high, which cannot be readily addressed by using a large antenna array. Instead, RISs can be deployed  to construct an auxiliary transmission link even when the direct link is blocked.

\subsection{RIS-aided Multicell Networks}

To maximize   spectrum efficiency (SE), multiple BSs in different cells reuse the same scarce frequency resources, which leads to   inter-cell interference, especially for the cell-edge users. Specifically, the  desired signal power received by the cell-edge user from its serving BS is comparable to  the interference received from its neighbouring cells. Hence, the cell-edge users  suffer from a low signal-to-interference-plus-noise ratio (SINR). To address this issue, the authors of \cite{panmulticell}   proposed to deploy an RIS at the cell boundary as shown in Fig. \ref{varioussystems}. In such a setting, the RIS is able to simultaneously enhance the signal gleaned from the serving BS, and mitigate the interference from the other. The simulation results of \cite{panmulticell} showed that the   sum rate achieved by an RIS-aided system having 80 reflecting elements may double that without an RIS.

\subsection{RIS-aided Simultaneous Wireless Information and Power Transfer  (SWIPT) Networks}
SWIPT is a promising technique of providing cost-effective power delivery to energy-limited internet of things (IoT) networks, where a BS with constant power supply broadcasts wireless signal to   information receivers (IRs) and energy receivers (ERs) simultaneously. The key challenge in SWIPT systems is that the ERs and IRs operate under different power supply requirements. Explicitly, ERs require  a received power on the order much higher than IRs. As a result, ERs should be deployed in closer proximity to the BS than IRs to harvest sufficient power, since the signal attenuation limits the ERs' practical operational range. To deal with this problem, the authors of \cite{panjsac} proposed to deploy an RIS in the proximity of the ERs, as shown in Fig. \ref{varioussystems}.  The simulation results of \cite{panjsac} revealed that  to ensure a minimum harvested power of 0.2 mW, the operational range of the ERs can be extended from 5.5 meters to 9 meters, when the RIS is equipped with  40 reflecting elements.

\subsection{RIS-aided Mobile Edge Computing  (MEC) Networks}

In  novel future applications such as virtual reality (VR), computation-intensive   image and video processing tasks must be executed in real time.  However, due to the limited power supply and hardware capabilities of typical VR devices,   these tasks cannot be accomplished locally. To tackle this issue, these computationally intensive tasks can be offloaded to   powerful computing nodes that are usually deployed at the edge of the network.  However, for some special cases where these devices are far from the MEC node, they can suffer from a low data offloading rate due to the severe path loss, which leads to excessive offloading delays. To overcome this impediment, a novel RIS-aided MEC framework was   proposed in \cite{tongbaijsac}, as shown in Fig. \ref{varioussystems}. The simulation results of \cite{tongbaijsac} showed that the overall task latency can be reduced from 115 ms to 65ms, if a 100-element RIS is employed.
\subsection{RIS-aided Non-orthogonal Multiple Access (NOMA)} 
NOMA constitutes a promising future multiple access technique in future wireless networks, in which each orthogonal resource block is shared by multiple simultaneous users. This significantly enhances the spectral efficiency of conventional orthogonal multiple access (OMA). However, in some special cases when the users' channel vectors are orthogonal to each other, NOMA may not be a good option. The ideal implementation scenario for NOMA is when all the users' channel vectors represent   the same angular direction. To broaden the application of NOMA, RIS can be introduced into the system for beneficially manipulating the wireless channel vectors of all users,  so that one user's channel vector can be aligned with the other one's \cite{zhiguo2020}.
\subsection{RIS-aided Multicast Networks}
Multicast transmission based on content reuse has attracted wide research attention, since it is capable of mitigating the tele-traffic, hence it will play a pivotal role in future wireless networks. Some typical examples using multicast transmission include video conferencing, video gaming and TV broadcast. In multi-group multicast communications,  identical content  is shared within each group, and each group's data rate is limited by the user with the weakest channel gain. To deal with this issue, an RIS-aided multicast architecture  was  proposed in \cite{guizhou2020} as shown in Fig. \ref{varioussystems}. By carefully tuning the RIS phase shifts, the channel conditions of the  weakest link can be enhanced.

\subsection{RIS-aided Physical Layer Security (PLS) Networks}
Due to the broadcast nature of wireless transmission, wireless links are prone to security threats such as jamming attacks or secure information leakage.  Recently, PLS techniques have received extensive research attention, since they can avoid  complex   key exchange protocols, and are suitable for latency-sensitive applications.   In order to maximize the rate of a secure communication link,   both   artificial noise and multiple  antennas have been proposed. However, when both the legitimate users and eavesdroppers have correlated channels or   when the    eavesdroppers are closer to the BS than the legitimate users, the achievable secure rate remains limited. To tackle this issue, in \cite{miaocui2019}, an RIS was deployed in a network operating in the presence of an eavesdropper as shown in Fig. \ref{varioussystems}, for mitigating the information leakage to the eavesdroppers, while simultaneously increasing the received signal power at the legitimate users.

\subsection{RIS-aided Cognitive Radio (CR) Networks}
CRs are capable of  enhancing the   SE by allowing secondary users (SUs) to reuse the same spectrum with primary users (PUs) while controlling the interference inflicted by the SU transmitters (SU-TXs) on the PU receivers (PU-RXs). A standard approach is to  use beamforming   for maximizing the sum-rate of the SUs, while ensuring that the interference power at the PU-RXs remains below the interference temperature (IT) limit. However, the   beamforming gain is limited, when the  SU-TX to SU-RX link is weak, and  the channel gain between the SU-TX and PU-RX is much higher. To handle this issue,  an RIS can be deployed in the vicinity of the PU-RXs, as shown in Fig. \ref{varioussystems}. The RIS is used for mitigating the interference towards the PU-RXs, while  improving the signal power at the SU-RXs.

\section{Case Study}
\begin{figure}
            \centering
              \includegraphics[width=0.5\textwidth]{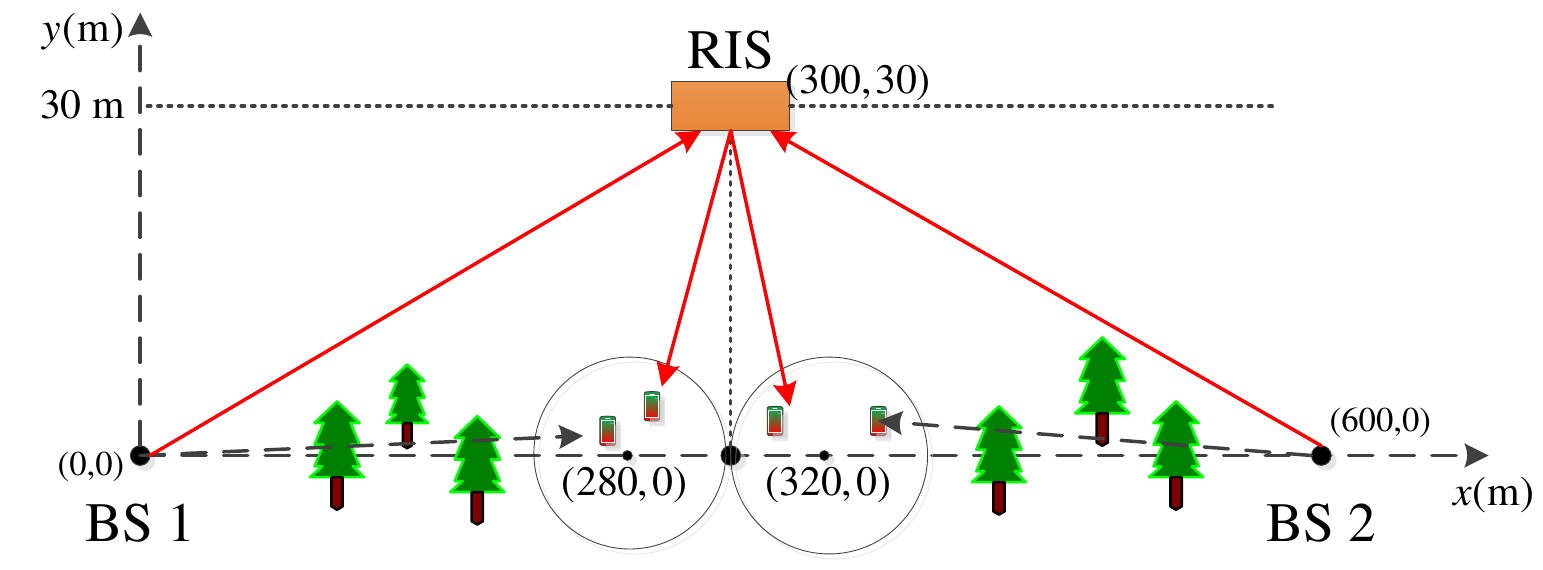}
              \caption{Simulation setup.}
              \label{simul}
 \end{figure}

In this section, we present a case-study to show the benefits of deploying an RIS in multicell networks. Specifically, we consider the RIS-aided two-cell network shown in Fig. \ref{simul}. Each cell has two users that are randomly positioned in a circle with radius of 20 m. The path-loss exponent between the BS and the users is  ${\alpha _{{\rm{BU}}}} = 3.75$. The path-loss exponents of the BS-RIS link
and of the RIS-user link are set to ${\alpha _{{\rm{BI}}}}={\alpha _{{\rm{IU}}}} \buildrel \Delta \over = {\alpha _{{\rm{RIS}}}}= 2.2 $. The small-scale fading is assumed to be Rayleigh fading. All other parameters are given in \cite{panmulticell}.

Explicitly, in \cite{panmulticell},    an iterative algorithm was proposed for jointly optimizing the transmit precoding at the BS and the passive beamforming at the RIS for maximizing the weighted sum rate (WSR). In the following, we compare it (termed as `RIS-aided') to the following two benchmark schemes:
\begin{enumerate}
  \item \textbf{RandPhase}: The phase shifts of the RIS are randomly generated from $[0,2\pi]$. The transmit precoding at the BS is designed based on the method of \cite{panmulticell}.
  \item \textbf{No-RIS}: There is no RIS in the system.
\end{enumerate}

 \begin{figure}
            \centering
              \includegraphics[width=0.4\textwidth]{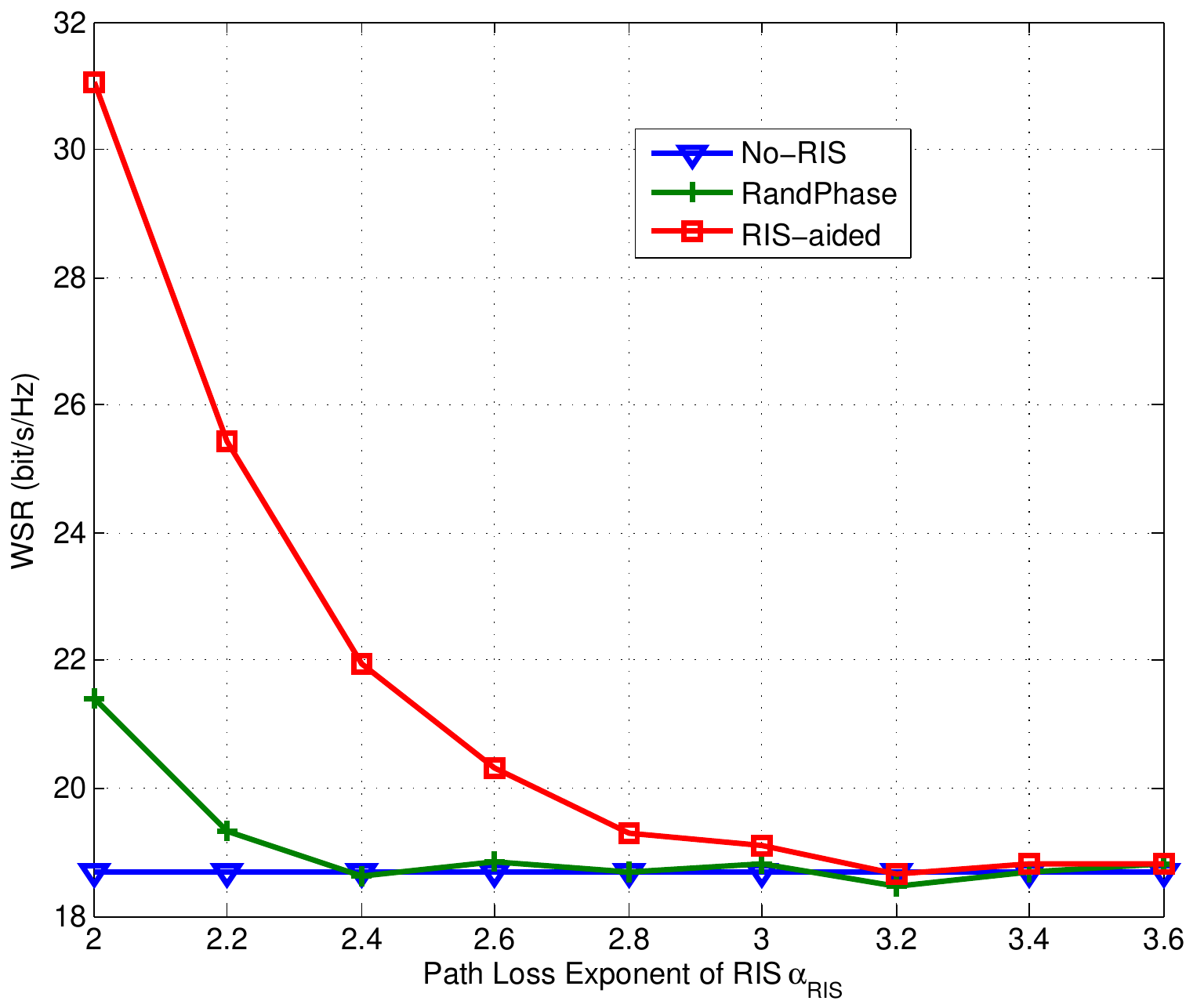}
              \caption{Weighted sum rate versus the path loss exponent ${\alpha _{{\rm{RIS}}}}$.}
              \label{vspathloss}
 \end{figure}

 \begin{figure}
            \centering
              \includegraphics[width=0.4\textwidth]{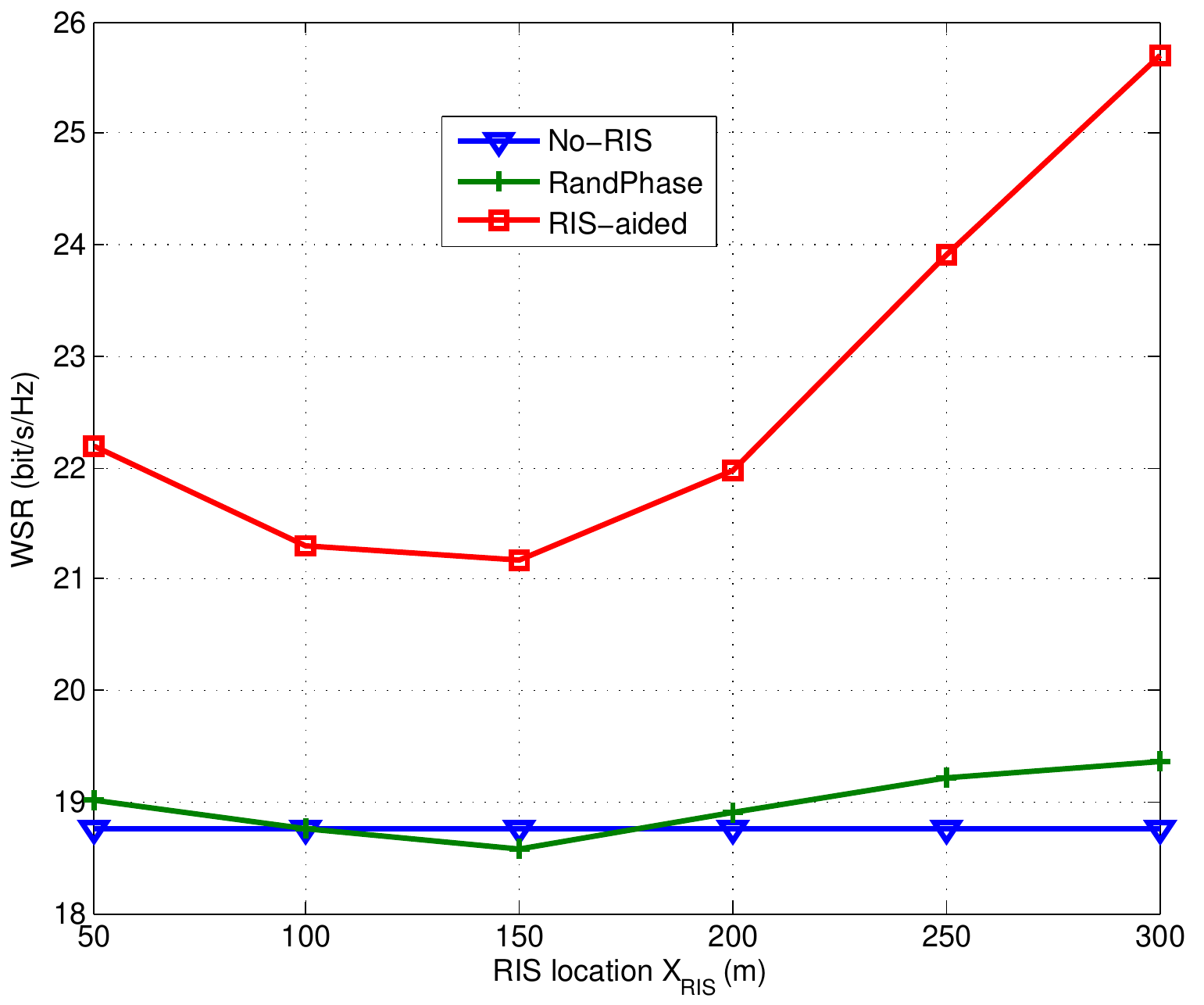}
              \caption{Weighted sum rate versus the RIS location $X_{\rm{RIS}}$.}
              \label{vsRISlocation}
 \end{figure}
Firstly, we study the impact of the RIS-related path loss exponent ${\alpha _{{\rm{RIS}}}}$ on the WSR performance in Fig.~\ref{vspathloss}. It can be observed from this figure that the WSR achieved by the `RIS-aided' algorithm decreases with the increase of ${\alpha _{{\rm{RIS}}}}$ and finally converges to the   WSR of the `no-RIS' scheme. The reason for this is that by increasing ${\alpha _{{\rm{RIS}}}}$, the RIS-related links suffer from severe signal attenuation, hence the signal reflected by the RIS becomes too weak. However, when ${\alpha _{{\rm{RIS}}}}$ is small, significant performance gains can be achieved by   an RIS   over the scenariooperating without RIS. This provides important   design insights indicating that the location of RIS should be carefully chosen for avoiding obstacles both in the BS-IRS link and IRS-user link.

Then, the impact of RIS location $X_{\rm{RIS}}$ on the WSR performance is evaluated in Fig. \ref{vsRISlocation}, where the RIS moves from $X_{\rm{IRS}}=50$ m (Cell center) to $X_{\rm{IRS}}=300$ m (cell edge). The `RIS-aided' algorithm is shown to achieve superior performance over the other two benchmark schemes. It is interesting to observe that the WSR obtained by the `RIS-aided' algorithm first decreases for $50$ m$ <X_{\rm{IRS}}<150$ m and then increases with $X_{\rm{IRS}}$. The minimum WSR value is achieved when the RIS is located mid-way between BS 1 and the RIS. This is in contrast to the conventional relaying scheme, where the best performance is achieved when the relay is halfway between the transmitter and the receiver. Furthermore, the WSR achieved when the RIS is  at the cell edge is much higher than in the case when the RIS is at the cell center. This is because positioning the RIS near BS 1 is only beneficial for users in cell 1, while all users benefit from the RIS, when  it is at the cell edge.

\section{What Are the Relevant Challenges and Research Directions?}
Although RISs are indeed appealing for the above applications, their implementation poses  several challenges as well. In the following, we list eight promising  future research directions.

\textbf{Direction 1: Channel Estimation}

To reap the full benefits  of RIS-aided wireless networks, the CSI should be estimated in support of the   phase shift design. We may consider near-instantaneous CSI and long-term CSI.

Concerning the near-instantaneous CSI estimation, we consider a typical RIS-aided wireless system  where a multi-antenna BS serves a single-antenna user with the aid of an RIS. Let us denote    the channel spanning from the BS to the RIS and that from the RIS to the user by  $\bf{H}$ and ${\bf{h}}_r$, respectively. In most situations having only  the cascaded CSI of the two hops defined as $\mathbf{G}=\mathrm{diag}(\mathbf{h}_{r}^{\mathrm{H}})\mathbf{H}$ is sufficient. However, since the number of reflecting elements is usually very large, the cascaded channel $\mathbf{G}$ contains a large number of channel coefficients. Hence, their estimation requires a large number of pilots, which is proportional to the number of reflecting elements. How to reduce the channel estimation overhead remains an open problem.

To facilitate the phase shift design based on long-term CSI, the angle or location information should be available at the BS. Unfortunately, conventional angle/location estimation algorithms may not be applicable for RIS-aided networks, when the direct channel spanning from the BS to the user is blocked. This is because a conventional BS can transmit pilot beams for tracking the users,  while the RIS is passive, hence cannot send pilot signals. Thus, low-complexity yet high-performance angle/location algorithms have to be conceived for RIS-aided networks.

\textbf{Direction 2: Passive Beamforming Design Based on Imperfect Parameters}

Given the estimated CSI, the phase shifts   have to be jointly designed together with the BS's active beamforming for achieving the desired   objectives. However, most of the existing contributions are based on the assumption of perfect instantaneous CSI, which is unrealistic in practice. For estimating the cascaded CSI, one should first estimate the direct BS-user channel by switching off the RIS, and then estimate the overall channel by switching on the RIS. The cascaded CSI can be calculated by subtracting the direct BS-user channel response from the overall channel response. Since the direct BS-user channel cannot be perfectly estimated, the subtraction operation will further contaminate the cascaded CSI, similar to the  error propagation of successive interference cancellation. Thus, the cascaded CSI error is sizeable and should be taken into account. An initial attempt was devoted to a simple RIS-aided multiple-user downlink scenario in \cite{guizhoutspframework}, and more emerging application scenarios are imperative to be investigated.

Furthermore, in addition to imperfect CSI, we encounter realistic transceiver hardware impairments (HWI), caused by non-linear amplifiers, low-resolution analog-to-digital converters (ADCs) and imperfect oscillators.   To reduce the hardware cost and power consumption,  a limited number of quantized phase-shifts may be used. This will impose quantization noise on the phase shifts of the RIS elements. Additionally, the popularly assumed phase-only reflection model is not  accurate in practice, since the reflection amplitude tends to depend on the value of the phase shift itself. Compared to  conventional  systems operating without RIS, the impact of HWIs in RIS-aided systems is complex due to the presence of quantized phase at the RIS. Hence, robust transmission design  is   needed that takes  into account  the  HWIs at both the transceivers and the RIS.

On the other hand, acquiring the near-instantaneous CSI may be challenging for the following reasons. Firstly, the training overhead is excessive when the number of reflecting elements is large. Secondly, the BS has to compute its beamforming weights as well as the phase shifts at the RIS, which entails solving a large-scale optimization problem. Thirdly, when the number of RIS elements is large and the channel is rapidly varying, the required capacity of the feedback link from the BS to the RIS also increases, which imposes a high overhead and high cost. To address these challenges, it is appealing to design the  phase shifts based on the long-term CSI relying on both angular and location information, which changes much more slowly. Unfortunately, only very few contributions were devoted to this research area \cite{yuhan}.

\textbf{Direction 3: Distributed Algorithms with Low Overhead Exchange}

In the RIS-aided multicell scenario of \cite{panmulticell}, the transmission design   is centralized. In particular, the  algorithm proposed requires a central processing unit (CPU) for collecting all the complex-valued channel matrices over the network. The CPU   computes all   the active beamforming weights and phase shifts, and then sends them back to the corresponding nodes.  However, these centralized algorithms suffer  from a heavy feedback overhead and high computational complexity, which is an impediment. Note that compared to conventional RIS-free systems, the large-dimensional cascaded channel matrix must additionally be fed back to the CPU.

Hence, it is imperative to design distributed algorithms, where each BS can make transmission decisions based on its local CSI and limited information exchange with other BSs.  Distributed algorithms have  appealing advantages over  centralized algorithms, such as low information exchange overhead, reduced computational complexity  and increased scalability.

\textbf{Direction 4: System Design for RIS-aided Frequency-division Duplex (FDD) systems}

Most  existing contributions related to RIS have considered   channel estimation for time-division duplex (TDD)-based implementations due to the appealing feature of channel reciprocity. However,   recent results in \cite{weicongchen} revealed that the RIS phase-shift model depends on the incident  electromagnetic angles, which implies that the assumption of channel reciprocity in TDD systems may not hold in practice. Hence, it is imperative to study  channel estimation and transmission design for FDD RIS systems. Due to a large number of reflecting elements at the RIS, large-dimensional channel matrices have to be fed back to the BS in FDD RIS systems, which incurs high feedback overhead.

\textbf{Direction 5: Application of RIS in Terahertz Communications}

Compared to mmWave communications, Terahertz (THz) communications can provide more abundant  bandwidth  and  higher data rates.   Despite its abundant bandwidth, THz frequencies suffer   from  strong atmospheric attenuation, molecular absorption and extremely severe path loss, which limits its operational range. Moreover, such high-frequency signals are very prone to blockage effects, and thus  they are unable to support reliable communication links. These drawbacks impair  its practical implementation. The use of RISs is a promising remedy to address these issues due to their ability to create alternative signal paths. In the THz band,   path-loss peaks appear in different frequency bands, and thus the total bandwidth has to be divided into several sub-bands having different bandwidths. Furthermore, THz electronic components can be made compact, and  thus the RIS can accommodate a massive number of tiny reflecting elements, allowing us to realize a holographic array having a   near-continuous aperture. Hence, channel estimation and beam pattern design for RIS-aided THz communications is an exciting area for future study.

%

%

\textbf{Direction 6: Mobility Management}

Mobility management is a challenging problem for RIS-aided wireless networks. Due to the rapid movements of   users, the BS may lose its connection  with them, unless agile mobility management schemes are used.  Since RISs are passive, they cannot send pilot signals to track the movement of the users. Hence, it is much more challenging to track roaming users, especially when the direct links between the BS and users are blocked.

\textbf{Direction 7: Deployment Issues}

The deployment strategy for RIS reflecting elements has a significant impact on the generation of the RIS-related channel coefficients and thus also on the system performance limit. From a practical implementation perspective, RIS deployments have to take into account the hardware cost, location availability, user distributions, and the requested services. There is a fundamental question to be answered: \emph{Given a total number of reflection elements, is centralized deployment better than distributed deployment?} To answer this question, we have to address the following two research issues:
\begin{enumerate}
  \item For centralized deployment, where to deploy the RIS? In the vicinity of the BS or the users, or where else between the BS and the user?
  \item For distributed deployment, how many smaller-size groups of RISs should the total number of reflecting elements be partitioned into? Where to deploy these smaller-size RISs?
\end{enumerate}

\textbf{Direction 8: AI-driven Design and Optimization}

The RIS phase shift matrices have to be optimized for enhancing the system performance. In practical deployments, each RIS is equipped with hundreds of reflecting elements. Most of the existing contributions on phase shift design tend to rely on model-based optimization methods, which require a large number of iterations to find a near-optimal solution, which is due to the non-convexity of phase shift constraints and owing to the non-convex nature of the objective function. Thus, the existing methods will incur high computational complexity, which are not suitable for real-time applications. To address this issue, artificial intelligence (AI)-based method is an appealing  data-driven scheme which can extract system features without a specific mathematical model. Once trained, the optimal solution can be found by simple algebraic calculations. Additionally, the trained model can be quite robust both against  imperfect CSI and hardware impairments.

\section{Conclusions}\label{hrforh}

In this article, we have answered four critical questions associated with RIS. Additionally, we have demonstrated that they are capable of mitigating the challenging blockage and coverage issues of mmWave or THz communications. We briefly introduced the basic RIS hardware architecture and its main advantages over  relays. We also discussed their potential integration into emerging wireless applications, including multicell, SWIPT, MEC, NOMA, multicast, PLS,   CR systems, mmWave. Finally, to provide  useful guidance and spark additional research interests, we also  formulated eight promising research directions.







\bibliographystyle{IEEEtran}
\bibliography{myre}
\textbf{ Cunhua Pan} received    Ph.D. degrees from  Southeast University, China, in 2015.  From 2015 to 2016, he was a Research Associate at the University of Kent, U.K. He held a post-doctoral position at Queen Mary University of London, U.K., from 2016 to 2019, where he is currently a Lecturer. He serves as Lead Guest Editor of IEEE JSTSP special issue on RIS,  Editor of IEEE CL, IEEE WCL, and IEEE ACCESS.

\textbf{Hong Ren} received    Ph.D. degrees from Southeast University,  China, in 2018. From   2016 to    2018, she was a visiting student in  University of Southampton, UK. She was   a postdoctoral scholar in  Queen Mary University of London, U.K., from 2018 to 2020. She is currently Associate Professor in Southeast University.

\textbf{Kezhi Wang} received    Ph.D. degree from   University of Warwick, U.K. in 2015. He was a Senior Research Officer in University of Essex, U.K. Currently he is a Senior Lecturer in   Northumbria University, U.K.

\textbf{Jonas Florentin Kolb} is now a PhD student in Queen Mary University of London, U.K.

\textbf{Maged Elkashlan} received   Ph.D. degree  from the University of British Columbia, Canada, 2006. From 2007 to 2011, he was with the CSIRO, Australia.  In 2011, he joined Queen Mary University of London, UK.

\textbf{Ming Chen} received   PhD degree from Nanjing University, in 1996. Since 1996, he joined Southeast University, where he is now a full professor.

\textbf{Marco Di Renzo} is a CNRS Research Director in the Laboratory of Signals and Systems of Paris-Saclay University, Paris, France. He serves as the Editor-in-Chief of IEEE Communications Letters. He is as IET Fellow (2020), an IEEE Fellow, and a Highly Cited Researcher (2019).

\textbf{Yang Hao}   received  Ph.D. degree in University of Bristol, U.K., in 1998.   He is currently a Professor of antennas
and electromagnetics with the Antenna Engineering Group, Queen Mary University of London, London, U.K.    He was a former EIC of the IEEE APC, and Editor of IEEE TAP.

\textbf{Jiangzhou Wang}  is  a Professor at the University of Kent, U.K.  He was the TPC Chair of IEEE ICC2019. He was the Executive Chair of IEEE ICC2015 and the TPC Chair of IEEE WCNC2013.

\textbf{A. Lee Swindlehurst} received the  PhD (1991) degree in  Stanford University. He was with
  Brigham Young University (BYU) from 1990-2007.   Since 2007 he has been a Professor   at the University of California Irvine.   Dr. Swindlehurst   was the inaugural EIC of the IEEE JSTSP.

\textbf{Xiaohu You} received the  Ph.D. degrees   from
Southeast University, Nanjing, China, in   1988. Since 1990, he has
been working  at Southeast University.
Dr. You was a recipient of  the National
Technological Invention Award of China in 2011.

\textbf{Hanzo Lajos} received  his doctorate in 1983.  He holds an honorary doctorate by the Technical University of Budapest (2009) and by the University of Edinburgh (2015).  He is a member of the Hungarian Academy of Sciences and a former EIC of the IEEE Press. He is a Governor of both IEEE ComSoc and of VTS.

\end{document}